\documentclass[aps,pre,twocolumn,showpacs,preprintnumbers]{revtex4-1}

\usepackage{graphicx}  % Include figure files
\usepackage{subfigure}
\usepackage{multirow}

\usepackage{fancyhdr}
\usepackage{longtable}
\usepackage{parskip}
\usepackage[T1]{fontenc}
\usepackage{dcolumn}   % Align table columns on decimal point
\usepackage{natbib}
\usepackage{bm}        % bold math
\usepackage{amsfonts}  % Common math fonts
\usepackage{amsmath}   % Common math functions
\usepackage{amssymb}   % Common math symbols
\usepackage{bbold}
\usepackage{relsize}
\usepackage{url,hyperref}
 \usepackage[table,usenames,dvipsnames]{xcolor}
\hypersetup{colorlinks=true, linkcolor=BrickRed,
  urlcolor=blue!50!black, citecolor=blue!50!black}
\usepackage[capitalize]{cleveref}
\crefrangelabelformat{equation}{(#3#1#4)$-$(#5#2#6)}
\let\ref\cref
\setcitestyle{super} 

\newcommand{\pwisein}{\left\{ \begin{array}{ll}}
\newcommand{\pwiseout}{\end{array}\right.}

\renewcommand{\Vec}[1]{\mathbf{#1}}

\newcounter{myequation}
\makeatletter
\@addtoreset{equation}{myequation}
\makeatother

\newcounter{myfigure}
\makeatletter
\@addtoreset{figure}{myfigure}
\makeatother

\begin{document}

\title{Finite size effect in the persistence probability of the
  Edwards-Wilkinson model of surface growth and effect of non-linearity.}
\author{Anirban Ghosh}
\email{anirbansonapur@gmail.com}
\email{anirbaniiser@gmail.com}
\affiliation{Department of Physics, Amity Institute of Applied Sciences, Amity University, Kolkata}
\affiliation{Indian Institute of Science Education and Research, Mohali}
\affiliation{Raman Research Institute, Bengaluru}

\author{Dipanjan Chakraborty}
\email{chakraborty@iisermohali.ac.in}
\affiliation{Indian Institute of Science Education and Research, Mohali}

\date{\today}

\begin{abstract}  

  The dynamical evolution of the surface height is controlled by
  either a linear or a nonlinear Langevin equation, depending on the
  underlying microscopic dynamics, and is often done theoretically
  using stochastic coarse-grained growth equations. The persistence
  probability $p(t)$ of stochastic models of surface growth that are
  constrained by a finite system size is examined in this work. We
  focus on the linear Edwards-Wilkinson model (EW) and the nonlinear
  Kardar-Parisi-Zhang model, two specific models of surface
  growth. The persistence exponents in the continuum version of these
  two models have been widely investigated. Krug et al.[Phys. Rev. E ,
  56:2702-2712, (1997)] and Kallabis et al. [EPL (Europhysics Letters)
  , 45(1):20, 1999] had shown that, the steady-state persistence
  exponents for both these models are related to the growth exponent
  $\beta$ as $\theta=1-\beta$. It is numerically found that the values
  of persistence exponents for both these models are close to the
  analytically predicted values. While the results of the continuum
  equations of the surface growth are well known, we focus to study
  the persistence probability expressions for discrete models with a
  finite size effect. In this article, we have investigated the
  persistence probabilities for the linear Edwards-Wilkinson(EW) model
  and for the non-linear Kardar-Parisi-Zhang(KPZ) model of surface
  growth on a finite one-dimensional lattice. The interesting
  phenomenon which is found in this case is that the known scenario of
  $p(t)$ of the following algebraic decay vanishes as we introduce a
  finite system size.

\end{abstract}

\maketitle 

\section{Introduction}

Surface growth is a common phenomenon in many processes of fundamental
interest and applied fields, occurring over a broad range of length
scales -- with atomistic growth models that range from few nanometers
to biological systems (such as growth of tumours) that range to few
millimeters. \cite{barabasi_stanley_1995,wakita1997} Such depsoition
processes are inherently spatially extended systems that are
stochastic in nature. Theoretical modelling of such systems is usually
done using stochastic coarse-grained growth equations -- the dynamical
evolution of the surface height is governed by either a linear or a
non-linear Langevin equation, depending on the underlying microscopic
dynamics. Specifically, a fluctuating interface is expressed and
characterized by it's height $h(\Vec{r},t)$. This is a single-valued
time-dependent scalar field defined at each point $\Vec{r}$ of a $d$
dimensional substrate of linear extent $L$. The size of the
fluctuation at a certain time $t$ is a function of the system size $L$
and time $t$ and is quantified through
$W(L,t)=[\langle[h(\Vec{r},t)-\Bar{h}(t)]^2\rangle]^{1/2}$, where
$\Bar{h}(t)=(1/V)\int h(\Vec{r},t)d\Vec{r}$ is described as spatially
averaged height with substrate volume $V$. $W(L,t)$ has scaling
behavior $ W(L,t)\sim t^\beta$ when $ 0\ll t\ll L^z$ and
$W(L,t) \sim L^\alpha$ when $t\gg L^z$ Three exponents
$\alpha$(roughness exponent), $\beta$(growth exponent), and
$z$(dynamical exponent) characterize the universality class of the
interface. $t\ll L^z$ is growing regime where width grows ($\beta>0$)
and $t\gg L^z$ is the steady state regime when fluctuations become
time-independent. While the stationary dynamics of such processes are
quite straightforward, particulary for linear Langevin equations, the
non-stationary dynamics is more difficult measure. For the latter, the
concept of persistence becomes extremely useful. % The concept of
% persistence possesses an important role in describing random processes
% in this physical world.
The word "persistence" conveys the meaning of survival. Although the
concept of ``persistence'' is well known for over six decades
\cite{slepian1962}, its application in studying first-passage
properties in non-equilibrium systems is quite recent. Persistence is
quantified through the persistence probability $p(t)$ -the probability
that a stochastic variable has not changed its sign up to time $t$. In a
wide range of non-equilibrium systems $p(t)$ is found to decay
algebraically with an exponent $\theta$, so mathematically
$p(t)=t^{-\theta}$, where $\theta$ is a non-trivial exponent. This
algebraic decay has been well documented in various stochastic
non-equilibrium models such as random walk, Brownian motion and
diffusion processes\cite{
  majumdar1996,majumdar1998a,newman1998,Majumdar:1999tn,
  sire2000,bray2004a,chakraborty2007,chakraborty2008,chakraborty2009,
  chakraborty2012d,chakraborty2012c,takikawa2013,ghosh2020,ghosh2022persistence}, reaction-diffusion\cite{odonoghue2001,cantrell2020},
phase-ordered kinetics,fluctuating interfaces
\cite{krug1997,kallabis1999,majumdar2001a,constantin2004,majumdar2005a,singha2005,bray2013b,gupta2014},
critical dynamics\cite{majumdar1996b,chakraborty2007c,Henkel2009},
polymer dynamics\cite{bhattacharya2007}, financial
markets\cite{zheng2002,ren2005,constantin2005} and many
more\cite{Yurke1997,majumdar1998,
  jain1999,Wong:2001dr,menon2001,majumdar2001,dean2001a,menon2003,ray2004a,
  Soriano:2009br,Efraim:2011ks}.

In this article, we look at the persistence probability $p(t)$ of
stochastic models of surface growth which are restricted by a finite
system size. We look at two particular models of surface growth - the
linear Edwards-Wilkinson model (EW) and the effect of weak
nonlinearity in the EW model, akin to the Kardar-Parisi-Zhang model
(KPZ). The phenomenon of persistence in the continuum version of these
two models has been well studied and the persistence exponents are
also known.\cite{krug1997,kallabis1999} For instance, the steady-state
persistence exponents for both these models are related to the growth
exponent $\beta$ as $\theta=1-\beta$,\cite{krug1997,kallabis1999} even
though the KPZ equation is a nonlinear
equation.% Later numerical studies on the
% continuum Kardar-Parisi-Zhang model also showed that the steady state
% persistence exponent for non-Gaussian surface growth models also
% exhibited the same relation.\cite{kallabis1999}
 Numerically obtained values of the steady state persistence exponent
for the one-dimensional KPZ equation was found to be
$\theta\approx 0.66$, close to the predicted value of
$2/3$,\cite{kallabis1999} whereas for the EW model the exponent was
found to be $\approx 0.74$, close to the predicted value of $3/4$.
While these results are for continuum equations of surface growth,
expressions for the persistence probability in spatially discrete
surface growth models with finite size effects are not well known. Our
aim is to investigate the persistence probability for discrete models
of surface growth equations with a finite size. In an infinite
spatially extended system, the boundary conditions do not play a
significant role. The scenario changes when the system size is finite
and it is expected that the well known algebraic decay of $p(t)$ is
lost.

The rest of the paper is organized as follows: in \cref{II} we present
a brief introduction on the models of surface growth. In
~\cref{sec:persistence} we present our work on the persistence
probabilities for the Edwards-Wilkinson model of surface growth on a
finite one-dimensional lattice (~\cref{ssec:EW_persistence}) and for
the effect of weak nonlinearity in the EW model of surface growth on a
finite one-dimensional lattice (\cref{ssec:KPZ_persistence}).

\section{Models of stochastic Surface Growth}
\label{II}

The dynamic scaling behaviour of stochastic growth equations are
characterized into several universality classes. Every choice of
universality class is characterized by a set of scaling exponents
depending upon the dimensionality of the problem. The exponents are
denoted as $\alpha$, $\beta$ and $z$, when $\alpha$ represents
roughness exponent exploring the dependence of the amplitude of height
fluctuations in the steady state regime ($t>>L^z$) on the sample size
$L$, $\beta$ denotes the growth exponent describing the initial
power-law growth of the interface width in the transient regime
($1<<t<<L^z$), and $z$ represents the dynamical exponent related to
the system size dependence of the time when the interface width
attains saturation. We use the single-valued function $h(\bold{r},t)$
representing the height of the growing sample at position $\bold{r}$
and deposition time $t$.The interfacial height fluctuations are
denoted by the root-mean-squared height deviation which is interface
width, that is a function of the substrate size $L$ and deposition
time $t$:
\begin{equation}
    W(L,t)=\langle[h(\bold{r},t)-\Bar{h}(t)]^2\rangle^{1/2}
\end{equation}
here $\Bar{h}(t)$=average sample thickness. $W(L,t)\propto t^\beta$
for $t<<L^z$ and $W(L,t)\propto L^\alpha$ for $t>>L^z$, $L^z$ being
the equilibration time of the interface, when its stationary roughness
is fully developed. % Since it is convenient to write the evolution
% equations in terms of the deviation of the height from its spatial
% average value, $h(\bold{r},t)-\Bar{h}(t)$, from now on we will denote
% by $\Bar{h}(\bold{r},t)$ the interface height fluctuation measured
% from the average height.
Using different scaling exponents
$(\alpha, \beta, z)$ we get,

(a) The Edward-Wilkinson (EW) second-order linear equation: $\frac{1}{2}$, $\frac{1}{4}$, $2$

\begin{equation}
  \label{eq:EW_cont}
    \frac{\partial h(\bold{r},t)}{\partial t}=\nu\nabla^2h(\bold{r},t)+\eta(\bold{r},t)
\end{equation}

(b) The KPZ second-order nonlinear equation: $\frac{1}{2}$, $\frac{1}{3}$, $\frac{3}{2}$

\begin{equation}
  \label{eq:KPZ_cont}
    \frac{\partial h(\bold{r},t)}{\partial t}=\nu\nabla^2h(\bold{r},t)+\lambda|\nabla h(\bold{r},t)|^2+\eta(\bold{r},t)
\end{equation}

(c) The Mullins-Herring (MH) fourth-order linear equation: $\frac{3}{2}$, $\frac{3}{8}$, $4(1,\frac{1}{4},4)$

\begin{equation}
    \frac{\partial h(\bold{r},t)}{\partial t}=-\nu\nabla^4h(\bold{r},t)+\eta(\bold{r},t)
\end{equation}

(d) The MBE fourth-order nonlinear equation: $\frac{2}{3}$, $\frac{1}{5}$, $\frac{10}{3}$
\begin{equation}
    \frac{\partial h(\bold{r},t)}{\partial t}=-\nu\nabla^4h(\bold{r},t)+\lambda\nabla^2|(\nabla h(\bold{r},t))|^2+\eta(\bold{r},t)
\end{equation}

The term $\eta(\bold{r},t)$ represents the noise term. We assume that
the noise has Gaussian distribution with zero mean and correlator:
\begin{equation}
    \langle\eta(\bold{r_1},t_1)\eta(\bold{r_2},t_2)\rangle=D\delta(\bold{r_1}-\bold{r_2})\delta(t_1-t_2)
\end{equation}
\section{Calculation of Persistence}
\label{sec:persistence}
\subsection{Persistence for Edward-Wilkinson system on a finite
  lattice}
\label{ssec:EW_persistence}

We consider the Edwards-Wilkinson model of surface growth on a
one-dimensional lattice with a finite domain size extending from $-L$
to $L$. The finite domain is discretized into a one-dimensional
lattice with $2N$ points, such that $Na=L$, where the lattice spacing
is defined as $a$.  At each lattice point the height profile is
denoted as $h_n(t)$. The continuum stochastic model of surface growth
given in \cref{eq:EW_cont} in one dimension reads as
\begin{equation}
    \frac{\partial h(x,t)}{\partial t}=\nu\frac{\partial^2 h}{\partial x^2}+\eta(x,t)
    \label{EW}
\end{equation}
Where $\eta$ is the Gaussian stochastic noise. The correlations of the
$\eta$ are given by
\begin{equation}
  \label{eq:noise_corr}
  \begin{split}
  &\langle \eta(x,t) \rangle=0\\
  &\langle \eta(x,t) \eta(x',t') \rangle = 2 D \delta(t-t')\delta(x-x')
  \end{split}
\end{equation}
Here $D$ is the diffusion constant. The boundary condition is chosen to be
$\left.\frac{\partial h}{\partial x}\right)_{\pm L}=0$.  The
discretized form of \cref{EW} on a one dimensional lattice takes the
form
\begin{equation}
    \frac{\partial h_n(t)}{\partial t}=\frac{\nu}{a^2}\Big[h_{n+1}(t)+h_{n-1}(t)-2h_n\Big]+\frac{\eta_n}{\sqrt{a}}
    \label{EW_discrete}
\end{equation}
Note the $\sqrt{a}$ in \cref{EW_discrete} comes from the spatial delta
correlation of the noise in the continuum equation.  The formal
solution to \cref{EW_discrete} together with the boundary condition
for $h_n$ is given by
\begin{equation}
    h_n(t)=X_0+2\sum_{p}X_p\cos{k_pn}
    \label{sol}
  \end{equation}
  where $X_p$ are the Fourier modes for $p\neq 0$ and $X_0$ is the
  $p=0$ mode. The boundary condition dictates that $\sin{k_pN}=0$ and
  therfore we get $k_p=p\pi/N$, so that the formal solution takes the
  form
\begin{equation}
  h_n(t)=X_0+2\sum_{p}X_p\cos{\frac{p\pi n}{N}}
  \label{eq:formal_sol}
\end{equation}
Substituting \cref{eq:formal_sol} in \cref{EW_discrete}, we get for
$X_p$
\begin{equation}
  \label{eq:xp}
  \begin{split}
  \sum_p \dot{X}_p\cos \left(\frac{p \pi
      n}{N}\right)&=-\frac{2\nu}{a^2}\sum_p X_p\left[1-\cos \left(\frac{p
        \pi}{N}\right)\right]\cos\left(\frac{p\pi n}{N}\right)\\
  &+\frac{\eta_n}{2\sqrt{a}}
\end{split}
\end{equation}
We multiply throughout \cref{eq:xp} with $\cos q \pi n/N$ and carry
out a sum over $n$. The left hand side of \cref{eq:xp} gives
\onecolumngrid
\begin{equation}
  \label{eq:xp_rhs}
  \begin{split}
  \sum_p \sum_n\dot{X}_p\cos \left(\frac{p \pi
      n}{N}\right)\cos \left(\frac{q \pi
      n}{N}\right)=\sum_p \dot{X}_p \frac{1}{a}\int_{-L}^L\mathrm{d}x\cos \left(\frac{p \pi
      x}{L}\right)\cos \left(\frac{q \pi
      x}{L}\right)=\frac{L}{a}\sum_p \dot{X}_p \delta_{p,q}=\frac{L}{a}\dot{X}_q
  \end{split}
\end{equation}
\twocolumngrid
Similarly, the first term on the right hand side of \cref{eq:xp} becomes
\onecolumngrid
\begin{equation}
  \label{eq:xp_lhs_1}
  \begin{split}
  -\sum_p \sum_n k_p X_p\cos \left(\frac{p \pi
      n}{N}\right)\cos \left(\frac{q \pi
      n}{N}\right)=-\sum_pk_p X_p \frac{1}{a}\int_{-L}^L\mathrm{d}x\cos \left(\frac{p \pi
      x}{L}\right)\cos \left(\frac{q \pi
      x}{L}\right)=-\frac{L}{a}\sum_p X_p \delta_{p,q}=-\frac{L}{a}k_qX_q
  \end{split}
\end{equation}
\twocolumngrid
Where $k_p=\frac{2\nu}{a^2}(1-\cos{\frac{p\pi}{N}})$.In the limit of
$N \to \infty $ and $a\to 0$, such that $Na=L$ remains finite, we
approximate $k_p$ as $k_p=\frac{\nu p^2\pi^2}{L^2}$.
The equation for the time evolution of $X_p$ follows the stochastic
differential equation
\begin{equation}
    \frac{\partial X_p}{\partial t}=-k_pX_p+\eta_p
\label{eq:xp_fina}
 \end{equation}
 where the stochastic noise $\eta_p$ is given by
 \begin{equation}
   \label{eq:eta_p}
   \eta_p(t)=\frac{\sqrt{a}}{2L}\sum_n \eta_n \cos \left(\frac{p \pi n}{N}\right)
 \end{equation}
 The statistical correlations of $\eta_p$ follows from $\eta_n$. The
 first moment of $\eta_p$ is zero. The second moment is given by 
 \begin{equation}
   \label{eq:eta_p_corr}
   \begin{split}
 %  &\langle \eta_p \rangle=0 \\
   &\langle \eta_p(t)\eta_q(t')\rangle=\frac{a}{4L^2}\sum_{n,m} 
   \langle \eta_n(t)\eta_m(t')\rangle \cos \left(\frac{p \pi
       n}{N}\right)\cos \left(\frac{q \pi m}{N}\right)\\
   &=\frac{2D a}{4L^2}\delta(t-t')\sum_n\cos \left(\frac{p \pi
       n}{N}\right)\cos \left(\frac{q \pi n}{N}\right)\\
  &=\frac{2D a}{4L^2}\delta(t-t')\frac{L}{a}\delta_{p,q}=\frac{D}{2L}\delta_{p,q}\delta(t-t')
\end{split}
\end{equation}
The noise correlations for $p=0$ mode also follows from
\cref{eq:eta_p}. While the first moment remains zero due to the
Gaussian nature of $\eta_n$, the second moment is given by 
\begin{equation}
  \label{eq:eta_0_corr}
  \begin{split}
   \langle \eta_0(t)\eta_0(t')\rangle=\frac{a}{4L^2}\sum_{n,m} 
   &\langle \eta_n(t)\eta_m(t')\rangle =\frac{2D
     a}{4L^2}\delta(t-t')\sum_{n,m} \delta_{n,m}\\
  &=\frac{2D a}{4L^2}\delta(t-t')\sum_n =\frac{D}{L}\delta(t-t')
\end{split}
\end{equation}
In deriving the last line of \cref{eq:eta_0_corr}, we have used the
fact that $\sum_n=2N$ and $Na=L$. With the noise correlation at hand,
we now proceed to calculate the two-time correltion functions.
The solution for $X_p$, for $p \neq 0$,  is given by
\begin{equation}
  \label{eq:solution_XP}
  X_p(t)=\int_0^t \mathrm{d}t' e^{-k_p (t-t')}\eta_p(t')
\end{equation}
and for $p=0$, $X_0$ obeys the simple random walk equation
\begin{equation}
  \label{eq:xp_0}
  X_0(t)=\int_0^t \mathrm{d}t' \eta_0(t')
\end{equation}

With $t_1>t_2$, the two-time correlation function for $\langle X_p(t_1) X_q(t_2)
\rangle$ take the form
\begin{equation}
\label{eq:correlation_two_time_EW}
  \langle X_p(t_1) X_q(t_2)\rangle=
  \begin{cases}
    \begin{split}
    & \frac{D}{L} t_2  \quad \textrm{for $p=q=0$}\\
    &\frac{D}{2L} \delta_{p,q}\left(\frac{\nu
        \pi^2 p^2}{L^2}+\frac{\nu \pi^2 q^2}{L^2}\right)^{-1} \times\\
    &\Bigg[e^{-\frac{\nu \pi^2 p^2}{L^2} (t_1-t_2)}-e^{-\frac{\nu \pi^2
        p^2}{L^2}t_1} e^{-\frac{\nu \pi^2 q^2}{L^2}t_2}\Bigg]\\
    &\quad
    \textrm{for $p \neq q \neq 0$} \\
    \end{split}
  \end{cases}
\end{equation}

We now want to determine the persistence probability in such a
system. For this, we choose the height profile at $n=0$ as the
stochastic variable, corresponding to $x=0$ in the continumm
limit. Putting $n=0$ in the formal solution \cref{eq:formal_sol}, we get
\begin{equation}
    h_0(t)=X_0+2\sum_{p}X_p
\label{eq:h0}
\end{equation}
The two-time correlation function $\langle h_0(t_1) h_0(t_2) \rangle$
is given by 
\begin{equation}
    \langle h_0(t_1)h_0(t_2)\rangle=\langle X_0(t_1)X_0(t_2)\rangle+4\sum_{p,q}\langle X_p(t_1)X_q(t_2)\rangle
    \label{eq1}
  \end{equation}
Substituting the two-time correlation functions derived in
\cref{eq:correlation_two_time_EW}, and noting that the delta function
in \cref{eq:correlation_two_time_EW} for $p\neq q\neq0$ is removed by
the sum over $q$ in \cref{eq1} we get 
%Putting the values of $\langle X_0(t_1)X_0(t_2)\rangle$ and $\langle X_p(t_1)X_q(t_2)\rangle$ in Eq.(\ref{eq1}), we get
\begin{equation}
  \label{eq:h0_t1_t2}
    \begin{split}
      \langle
      h_0(t_1)h_0(t_2)\rangle&=\frac{D}{L}t_2+\frac{2D}{L}\sum_{p=1}^{\infty}\left(\frac{2\nu\pi^2
        p^2}{L^2}\right)^{-1}\times\\
      &\left[e^{-\frac{\nu\pi^2 p^2}{L^2}(t_1-t_2)}-
        e^{-\frac{\nu\pi^2 p^2}{L^2}(t_1+t_2)}\right]\\
%      &=2D\Big[t_2+\frac{L^2}{\nu\pi^2}e^{-\frac{\nu\pi^2}{L^2}t_1}\sinh{\frac{\nu\pi^2}{L^2}t_2}\Big]
    \end{split}
\end{equation}
With this expression in hand, we first consider the limit of $L \to
\infty$. To this end we use the Euler-Maclaurin formula for the sum
over the Fourier modes. 
\begin{equation}
  \label{eq:euler_maclaurin}
  \begin{split}
 & \sum_{p=1}^{\infty}\left(\frac{2\nu\pi^2
      p^2}{L^2}\right)^{-1}
  \left[e^{-\frac{\nu\pi^2 p^2}{L^2}(t_1-t_2)}-
    e^{-\frac{\nu\pi^2
        p^2}{L^2}(t_1+t_2)}\right]\\
  &=\frac{L}{2\pi}\int_0^\infty {e^{-\nu k^2 (t_1-t_2)}-e^{-\nu k^2
      (t_1+t_2)} \over \nu k^2}-\frac{1}{2}f(0)
\end{split}
\end{equation}
where $f(k)={e^{-\nu k^2 (t_1-t_2)}-e^{-\nu k^2
      (t_1+t_2)} \over  2\nu k^2}$
Therefore, in this limit of $L \to \infty$, the sum is rewritten as
\begin{equation}
    \begin{split}
        &\sum_{p=1}^{\infty}\frac{e^{-k_p(t_1-t_2)}-e^{-k_p(t_1+t_2)}}{2k_p}\\
        &=\int_{0}^{\infty}\frac{e^{-\nu k^2(t_1-t_2)}-e^{-\nu k^2(t_1+t_2)}}{2k_p}dk-\frac{1}{2}[f(0)+f(\infty)]
    \end{split}
    \label{33}
\end{equation}
Where $f(k_p)=\frac{e^{-k_p(t_1-t_2)}-e^{-k_p(t_1+t_2)}}{2k_p}$. In
the limit of $k \to 0$ we get $f(0)=t_2/2$ so that the expression in
\cref{eq:h0_t1_t2} becomes
\begin{equation}
  \label{eq:h0_t1_t2_L_infty}
    \begin{split}
      \langle
      h_0(t_1)h_0(t_2)\rangle&=\frac{D}{L}t_2+\frac{2D}{2\pi}\int_0^{\infty}\mathrm{d}k\;
      \frac{1}{\nu k^2}\left[e^{-\nu k^2(t_1-t_2)}\right.-\\
     & \qquad \qquad \qquad \qquad \qquad\left.  e^{-\nu k^2(t_1+t_2)} \right]
    -\frac{D}{L} t_2\\
    &=D\int_{-\infty}^{\infty}\frac{\mathrm{d}k}{2\pi}
      \left[{e^{-\nu k^2(t_1-t_2)}-
        e^{-\nu k^2(t_1+t_2)} \over \nu k^2}\right]
%      &=2D\Big[t_2+\frac{L^2}{\nu\pi^2}e^{-\frac{\nu\pi^2}{L^2}t_1}\sinh{\frac{\nu\pi^2}{L^2}t_2}\Big]
    \end{split}
\end{equation}
The final form of the two-time correlation function in
\cref{eq:h0_t1_t2_L_infty} is the well known result for the one
dimensional Edwards-Wilkinson model of surface growth in the continuum
limit.\cite{krug1997,constantin2004}
% Using the Eq.(\ref{33}) we find the two-time correlation function as
% \begin{equation}
%     \begin{split}
%     C(t_1,t_2)&=2D\int_{0}^{\infty}\frac{e^{-\nu k^2(t_1-t_2)}-e^{-\nu k^2(t_1+t_2)}}{2\nu k^2}dk
%     \end{split}
%     \label{34}
%   \end{equation}
Denoting $C(t_1,t_2) \equiv \langle h_0(t_1) h_0(t_2) \rangle$ we get
\begin{equation}
   C(t_1,t_2)=\frac{D}{\nu}[(t_1+t_2)^{1/2}-(t_1-t_2)^{1/2}]
 \end{equation}
 We now define the nromalised variable
 $H(t)=h_0(t)/\sqrt{\langle h_0^2(t) \rangle}$ and the two-time
   correlation function of $H(t)$,
   $A(t_1,t_2)\equiv \langle H(t_1)H(t_2) \rangle =
   C(t_1,t_2)/\sqrt{C(t_1,t_1)C(t_2,t_2)}$ is given by
\begin{equation}
    \begin{split}
      A(t_1,t_2)&=\frac{C(t_1,t_2)}{\sqrt{C(t_1,t_1)C(t_2,t_2)}}\\
      &=\Big[\frac{1}{2}\Big(\sqrt{\frac{t_1}{t_2}}+\sqrt{\frac{t_2}{t_1}}\Big)\Big]^{1/2}-\Big[\frac{1}{2}
      \Big(\sqrt{\frac{t_1}{t_2}}-\sqrt{\frac{t_2}{t_1}}\Big)\Big]^{1/2}
    \end{split}
\label{eq:A_t1_t2}
  \end{equation}
% This is Edward-Wilkinson interface
% results.\cite{krug1997persistence,constantin2004persistence}
The non-stationary correlation function in \cref{eq:A_t1_t2} is
transformed in to a stationary correlator using the transformation
$T=\ln t$, so that we get 
% Now we define logarithmic time variable $T_1=ln(t_1)$ and $T_2=ln(t_2)$. When
% the process is measured in logarithmic time scale, the process becomes
% stationary which depends on the time difference $T=T_1-T_2$
\begin{equation}
    A(T_1,T_2) =A(T_1-T_2)=f_0(T)=[\cosh{T/2}]^{1/2}-[\sinh{T/2}]^{1/2}
\end{equation}

% Using the formal solution in \cref{sol} 
% The correlation is expressed as

% The first term of the Eq.(\ref{eq1}) is 
% \begin{equation}
%     \langle X_0(t_1)X_0(t_2)\rangle=2Dt_2
% \end{equation}
% when $t_2<t_1$.
% Now Eq.(\ref{EW}) can be written as
% \begin{equation}
%     \frac{\partial h_n}{\partial t}=\frac{\nu}{a^2}[h_{n+1}+h_{n-1}-2h_n]+\xi_p
% \end{equation}

In the opposite limit of $L \to 0$, only the first term in
\cref{eq:h0_t1_t2} survives and the correlation function
for the normalised variable $H(t)=h_0(t)/\sqrt{\langle h_0^2(t) \rangle}$
is given by
\begin{equation}
  \label{eq:L_to_0}
  \langle H(t_1) H(t_2) \rangle=\sqrt{\frac{t_2}{t_1}}
\end{equation}
This is the result for a simple random walk and the non-stationary
correlation function is converted to a stationary correlator using the
transformation $T=\ln t$. In the imaginary time $T$, the two-time
correlation function becomes stationary:$\langle H(T_1)
H(T_2)\rangle=e^{-(T_1-T_2)/2}$ and the persistence probability is
that of a simple random walker $p(t) \sim t^{-1/2}$.

We now study the case when $L$ is finite.Thus, $L$ is kept fixed while
$t$ is varied. To this end, in the expression for the two-time
correlation function in \cref{eq:h0_t1_t2} we keep the long wavelength
mode $\pi/L$ corresponding to $p=1$. The two-time correlation function
becomes
\begin{equation}
  \label{eq:h0_t1_t2_finite_L}
    \begin{split}
      \langle
      h_0(t_1)h_0(t_2)\rangle=\frac{D}{L}\Big[t_2+\frac{2L^2}{\nu\pi^2}
      e^{-\frac{\nu\pi^2}{L^2}t_1}\sinh{\frac{\nu\pi^2}{L^2}t_2}\Big]
    \end{split} 
\end{equation}
The first limiting case we note is that of $\nu \to 0$ when each
lattice is independent of its neighboring site and evolves according
to a simple random walk model. In which case, we note that the
correlation function in \cref{eq:h0_t1_t2_finite_L} becomes that of a
simple random walker and consequently we expect the persistence
probability to be $p(t) \sim t^{-1/2}$. In order to proceed further,
we note that the non-stationary correlation function in
\cref{eq:h0_t1_t2_finite_L} in its exact form can not be transformed
to a stationary correlator without any further approximation. % We

When $t_1$ and $t_2$ are such that $\nu \pi^2 t/L^2 >>1$,
  the first term in \cref{eq:h0_t1_t2_finite_L} dominates and
  consequently the persistence probability is that of a random walker:
  $p(t) \sim t^{-1/2}$. In the opposite limit of $\nu \pi^2 t/L^2 <<1$
  we approximate the correlator in \cref{eq:h0_t1_t2_finite_L} as:
  \begin{equation}
    \label{eq:h0_t1_t2_finite_L_approx}
    \begin{split}
      \langle
      h_0(t_1)h_0(t_2)\rangle&=\frac{D}{L} t_2 \left[1+2
        \left(1-\frac{\nu\pi^2}{L^2}t_1\right)\frac{\sinh{\frac{\nu\pi^2}{L^2}t_2}}{\frac{\nu\pi^2}{L^2}t_2}\right]\\
      &=\frac{D}{L} t_2 \left[1+2
        \left(1-\frac{\nu\pi^2}{L^2}t_1\right)\right]\\
      &=\frac{3D}{L} t_2 \left[1-\frac{2}{3}\frac{\nu\pi^2}{L^2}t_1\right]
      \end{split}
\end{equation}

% We
% therefore consider the case when the values of $t_1$ and $t_2$ are such
% that they obey the inequality $L^2/\nu\pi^2t >> 1$. In this scenario,
% ignoring the first term within the bracket in
% \cref{eq:h0_t1_t2_finite_L}, we get
% \begin{equation}
%   \label{eq:h0_t1_t2_finite_L_approx}
%       \langle
%       h_0(t_1)h_0(t_2)\rangle=\frac{2DL}{\nu\pi^2}
%       e^{-\frac{\nu\pi^2}{L^2}t_1}\sinh{\frac{\nu\pi^2}{L^2}t_2}
% \end{equation}
We can now convert this to a stationary correlation function -- first using the
transformations $H(t)=h_0(t)/\sqrt{\langle h_0^2(t) \rangle}$ so that 
\begin{equation}
  \label{eq:H_t1_t2_approx_2}
        \langle
      H(t_1)H(t_2)\rangle= \sqrt{\frac{t_2}{t_1}}\left[\frac{1-(2/3){\nu \pi
       ^2 t_1 \over L^2}}{1-(2/3){\nu \pi^2 t_2 \over L^2}}\right]
      % \frac{{\scriptstyle (L^2/\nu\pi^2)} e^{-\frac{\nu\pi^2}{2L^2}t_1}}{{\scriptstyle
      %     (L^2/\nu\pi^2)}e^{-\frac{\nu\pi^2}{2L^2}t_2}}\left[\frac{\sinh{\frac{\nu\pi^2}{L^2}t_2}}{\sinh{\frac{\nu\pi^2}{L^2}t_1}}\right]^{1/2}
\end{equation}
\begin{figure}
    \centering
    \includegraphics[width=0.5\textwidth]{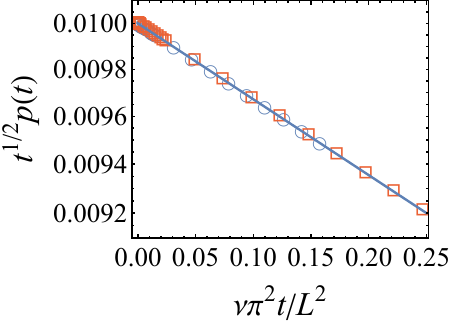}
    \caption{Semilog plot of $p(t)$ with $\nu\pi^2t/L^2$ of
      Edwards-Wilkinson surface growth in finite size lattice for
      different values of $a$ as written in the legends. Here $N=50$
      and $\nu=0.1$ values are kept fixed. The solid lines represent
      the analytical results from the expression of
      Eq.(\ref{eq:per_prob_1}).}
    \label{fig1}
\end{figure}

The transformation to a time $T$ is given by
\begin{equation}
  \label{eq:time_transformation}
  e^{T/2}={t^{1/2} \over \left(1-\frac{2}{3}\frac{\nu \pi^2 t}{L^2}\right)^{1/2}}
  %\frac{L^2}{\nu \pi^2} e^{\frac{\nu \pi^2 t}{L^2}}\sinh\frac{\nu \pi^2 t}{L^2}
\end{equation}
so that $\langle H(T_1)H(T_2)\rangle = e^{-(T_1-T_2)/2}$, and
following Slepian\cite{slepian1962}, the persistence probability in
real-time is given by
\begin{equation}
  \label{eq:per_prob}
  p(t) \sim \frac{1}{\sqrt{t}}\left(1-\frac{2}{3}\frac{\nu \pi^2 t}{L^2}\right)^{1/2}
  % \sqrt{\frac{\nu \pi^2}{L^2}}\frac{e^{-\frac{\nu \pi^2 t}{ 2L^2}}}{\sqrt{\sinh {\nu \pi^2 t
      % \over L^2}}}
\end{equation}
For term in the bracket in \cref{eq:per_prob} can be exponentiated to
get an alternate form for $p(t)$:
\begin{equation}
  \label{eq:per_prob_1}
  p(t) \sim \frac{1}{\sqrt{t}}e^{-\frac{1}{3}\frac{\nu \pi^2 t}{L^2}}
  % \sqrt{\frac{\nu \pi^2}{L^2}}\frac{e^{-\frac{\nu \pi^2 t}{ 2L^2}}}{\sqrt{\sinh {\nu \pi^2 t
      % \over L^2}}}
\end{equation}

Thus, the quantity $t^{1/2}p(t)$ decays exponentially as
$e^{-\nu \pi^2 t/3L^2}$ and the value of $\nu$ can be determined from
this decay.

The simulation results displayed in Fig.(\ref{fig1}) validate the
theoretical and simulated persistence probabilities for the
Edward-Wilkinson surface growth model, particularly concerning the
finite-size effect under the condition $\frac{\nu\pi^2t}{L^2}<<
1$. Fig.(\ref{fig1}) itself is a semilog plot of $t^{1/2}p(t)$ derived
from reorganizing Eq.(\ref{eq:per_prob_1}) versus $\nu\pi^2t/L^2$.

The simulation has been done using the dicrete version of EW eqution
in Eq.(\ref{EW_discrete}). The trajectories were evolved in time with
an integration time step of $\delta t=0.001$.
At each moment in time, the survival of a particle's trajectory was assessed by examining the value of 
$h(t)$. The survival probability 
$p(t)$ was determined as the fraction of trajectories where 
$h(t)$ had changed sign up to time $t$.
% At every instant, the
% survival of the particle trajectory was checked by looking at the
% value of $h(t)$. A fraction of trajectories for which $h(t)$ did not
% come below $h(0)$ up to time $t$ gave the survival probability
% $p(t)$.
A total of $10^8$ trajectories were used in estimating the
survival probability.

\subsection{The effect of non-linearity}
\label{ssec:KPZ_persistence}

\begin{figure}[!t]
    \centering
    \includegraphics[width=0.4\textwidth]{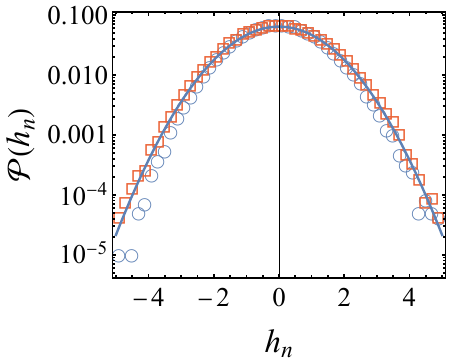}
    \caption{Probability distribution function $\mathcal{P}(h_n)$ in semilog scale at
      two different times $t=0.1$ (open circles) and $t=10$ (open
      squares) for total lattice sites
      $N=5$, lattice spacing $a=0.02$, $\nu=0.1$, and $\lambda=0.01$.}
    \label{fig:prob_dist_kpz}
\end{figure}

We now focus on the effect of non-linearity in the Edwards-Wilkinson
model. The non-linear term is of the form $\lambda (\nabla h)^2$,
where $\lambda$ the coupling constant. % We use discretized Kardar-Parisi-Zhang model of surface
% growth.
The conitnumm model in \cref{eq:KPZ_cont} in one-dimension
takes the form
\begin{equation}
    \frac{\partial h_n}{\partial t}=\nu\frac{\partial^2h_n}{\partial x^2}+\lambda\left(\frac{\partial h_n}{\partial x}\right)^2+\eta_n
 \label{7}
\end{equation}
The boundary conditions remain the same as in the preceeding section,
that is, $\left.\frac{\partial h}{\partial t}\right)_{\pm L}=0$. As
before, we spatially discretize the equation on a one-dimensional
lattice with a lattice spacing $a$:
\begin{equation}
    \frac{\partial h_n}{\partial
      t}=\frac{\nu}{a^2}[h_{n+1}+h_{n-1}-2h_n]+
    \lambda\left(\frac{h_{n+1}-h_{n-1}}{2a}\right)^2+\frac{\eta_n}{\sqrt{a}}
    \label{8}
\end{equation}
Here $\lambda$ is the non-linear coupling parameter. We choose a weak
$\lambda$ for two reasons- first a perturbative expansion around
$\lambda=0$ can be done and the solution to \cref{8} can be
constructed using the perturbative solution. Secondly, the choice of a
weak $\lambda$ is dictated by the requirement of $h(x,t)$ to be a
Gaussian process (see \cref{fig:prob_dist_kpz}). In the
Edwards-Wilkinson model since $h(x,t)$ is linear, the process remains
a Gaussian stochastic process. However, this is not true for \cref{7}
since it contains a nonlinear term. While this puts sever restrictions
on the study, nevertheless, for weak coupling the persistence
probability provides a way to measure the ratio $\lambda/\nu$.

We consider the perturbative expansion
\begin{equation}
  \label{eq:perturbative_expansion}
  h_n=h_n^0+\lambda h_n^1+\lambda^2h_n^2+ \mathcal{O}(\lambda^3)
\end{equation}
Substituting \cref{eq:perturbative_expansion} in \cref{8} and
comparing the left hand and the right side to the powers of $\lambda$
we get for $h_n^0$
\begin{equation}
    \frac{\partial h_n^0}{\partial t}=\frac{\nu}{a^2}[h_{n+1}^0+h_{n-1}^0-2h_n^0]+\frac{\eta_n}{\sqrt{a}}
    \label{9}
\end{equation}
and for $h_n^1$ as
\begin{equation}
    \frac{\partial h_n^1}{\partial
      t}=\frac{\nu}{a^2}[h_{n+1}^1+h_{n-1}^1-2h_n^1]+
    \Big(\frac{h_{n+1}^0-h_{n-1}^0}{2a}\Big)^2
    \label{20}
  \end{equation}
  To proceed further, we note that the solution given in
  \cref{eq:perturbative_expansion}, must obey the boundary condition
  at each order of $\lambda$. Specifically, one has
  $\left.\frac{\partial h_n^0}{\partial t}\right)_{\pm L}=0$,
  $\left.\frac{\partial h_n^1}{\partial t}\right)_{\pm L}=0$ and so
  on. Consequently, we write the solution as
  
% \begin{equation}
%     \begin{split}
%         &h_n=X_0+2\sum_{p=1}^{\infty}X_p\cos{\frac{p\pi n}{N}}\\
%         &=(X_0^0+\lambda X_0^1+\lambda^2X_0^2+.)+2\sum_{p=1}^{\infty}(X_p^0+\lambda X_p^1+\lambda^2X_p^2+.)\cos{\frac{p\pi n}{N}}
%     \end{split}
%     \label{10}
% \end{equation}
% Equalizing the coefficients of the terms in both sides
\begin{equation}
    \begin{split}
        &h_n^0=X_0^0+2\sum_{p}X_p^0\cos{\frac{p\pi n}{N}}\\
        &h_n^1=2\sum_{p}X_p^1\cos{\frac{p\pi n}{N}}\\
        &h_n^2=2\sum_{p}X_p^2\cos{\frac{p\pi n}{N}}
    \end{split}
    \label{11}
  \end{equation}
As before, we will be interested in the two-time correlation for
$h_0$. Using \cref{11}, the two-time correlation function $\langle
h_0(t_1)h_0(t_2)\rangle$ takes the form 
\begin{equation}
    \begin{split}
    \langle h_0(t_1)h_0(t_2)\rangle&=\langle h_0^0(t_1)h_0^0(t_2)\rangle+\lambda^2\langle h_0^1(t_1)h_0^1(t_2)\rangle\\
    &=\langle X_0^0(t_1)X_0^0(t_2)\rangle+4\sum_{p,q}\langle X_p^0(t_1)X_q^0(t_2)\rangle\\
    &+4\lambda^2\sum_{p,q}\langle X_p^1(t_1)X_q^1(t_2)\rangle
    \end{split}
    \label{19}
\end{equation}
In writing \cref{19}, we have ignored the term $\langle
h_0^0(t_1)h_0^2(t_2)\rangle$ and $\langle h_0^2(t_1)h_0^0(t_2)\rangle$
in the order $\lambda^2$ term since they contain higher order
exponential decays.

The equation for the zeroth order $h_n^0$ obeys the same differential
equation as that of the discrete Edwards-Wilkinson model (see
\cref{EW_discrete}) and therefore the solution for $h_n^0$ is
known. Consequently, the two-time correlation function $\langle X^0_p(t_1)
X^0_q(t_2) \rangle$ follows from \cref{eq:correlation_two_time_EW}.  

We focus on the solution of $h_n^1$. Substituting the expression of
$h_n^1$ in terms of $X_p^1$ from Eq.(\ref{11}), we get
\begin{equation}
    \begin{split}
      &\frac{\partial}{\partial t}\Big[2\sum_{p}X_p^1\cos{\frac{p\pi
          n}{N}}\Big]=\frac{\nu}{a^2}
      \Bigg[2\sum_{p}X_p^1\Big(\cos{\frac{p\pi (n+1)}{N}}\\
      &+\cos{\frac{p\pi (n-1)}{N}}-2\cos{\frac{p\pi
          n}{N}}\Big)\Bigg]+\frac{1}{4a^2}\Bigg[\sum_{p}X_p^0
      \cos{\frac{p\pi(n+1)}{N}}\\
      &-\sum_{p}X_p^0\cos{\frac{p\pi(n-1)}{N}}\Bigg]^2
    \end{split}
    \label{21}
\end{equation}
This equation can be simplified to 
\begin{equation}
    \begin{split}
      &\sum_{p}\frac{\partial X_p^1}{\partial t}\cos{\frac{p\pi
          n}{N}}=-\frac{2\nu}{a^2}\sum_{p}X_p^1
      \cos{\frac{p\pi n}{N}}\Big(1-\cos{\frac{p\pi }{N}}\Big)\\
      &+\frac{1}{2a^2}\sum_{p}X_p^0X_q^0\sin{\frac{p\pi
          n}{N}}\sin{\frac{p\pi}{N}}\sin{\frac{q\pi
          n}{N}}\sin{\frac{q\pi}{N}}
    \end{split}
    \label{22}
\end{equation}
Multiplying both sides with the factor $\cos{\frac{k\pi n}{N}}$ and
summing over $n$, we get
\begin{equation}
    \begin{split}
        &\sum_{p}\sum_n\frac{\partial X_p^1}{\partial t}\cos{\frac{p\pi n}{N}}\cos{\frac{k\pi n}{N}}\\
        &=-\frac{2\nu}{a^2}\sum_{p}\sum_nX_p^1\cos{\frac{p\pi
            n}{N}}\cos{\frac{k\pi n}{N}}
        \Big(1-\cos{\frac{p\pi}{N}}\Big)\\
        &+\frac{1}{2a^2}\sum_{p}\sum_nX_p^0X_q^0\sin{\frac{p\pi
            n}{N}}\sin{\frac{q\pi n}{N}}
        \cos{\frac{k\pi n}{N}}\sin{\frac{p\pi }{N}}\sin{\frac{q\pi}{N}}
    \end{split}
    \label{23}
\end{equation}
The term on the left hand side of \cref{23} and the first term on the
right hand side is easy to evaluate. The non-linear term on the right
hand side of \cref{23} takes the form

%\vspace{5cm}
%\textcolor{red}{The term of L.H.S. of Eq.(\ref{23}) can be expressed as}

% \onecolumngrid

% \textcolor{red}{\begin{equation}
% \sum_{n}\sum_{p}\dot{X}_{p}^1\cos{\frac{p\pi n}{N}}\cos{\frac{k\pi n}{N}}=\sum_{p}\dot{X}_{p}^1\frac{1}{a}\int_{-L}^{+L}dx\cos{\frac{p\pi x}{L}}\cos{\frac{k\pi x}{L}}=\frac{L}{a}\sum_{p}\dot{X}_{p}^1\delta_{p,k}=\frac{L}{a}\dot{X}_{k}^1
% \label{23a}
% \end{equation}}
%\twocolumngrid

% \textcolor{red}{The first term of R.H.S. of Eq.(\ref{23}) can be written as}
% \onecolumngrid
% \textcolor{red}{\begin{equation}
% -\sum_{p}\sum_{n}k_pX_p\cos{\frac{p\pi n}{N}}\cos{\frac{k\pi n}{N}}=-\sum_{p}k_pX_p\frac{1}{a}\int_{-L}^{+L}dx\cos{\frac{p\pi x}{L}}\cos{\frac{k\pi x}{L}}=-\sum_{p}k_pX_p\frac{L}{a}\delta_{p,k}=-\frac{L}{a}k_kX_k
% \label{23b}
% \end{equation}}

% \twocolumngrid

%\onecolumngrid
\begin{equation}
\begin{split}
&\frac{1}{2a^2}\sum_{p,q}\sum_nX_p^0X_q^0\sin{\frac{p\pi n}{N}}\sin{\frac{q\pi n}{N}}\cos{\frac{k\pi n}{N}}\sin{\frac{p\pi }{N}}\sin{\frac{q\pi}{N}}\\
&=\frac{1}{2a^2}\sum_{p,q}X_p^0X_q^0\sin{\frac{p\pi
  }{N}}\sin{\frac{q\pi}{N}}\times\\
&\qquad \qquad \qquad \qquad\qquad  \frac{1}{a}\int_{-L}^{+L}dx\sin{\frac{p\pi x}{L}}\cos{\frac{k\pi x}{L}}\sin{\frac{q\pi x}{L}}\\
&=\frac{1}{2a^2}\sum_{p,q}X_p^0X_q^0\sin{\frac{p\pi }{N}}\sin{\frac{q\pi}{N}}\frac{L}{2a}(\delta_{p,q-k}+\delta_{p,q+k})\\
&=\frac{L}{4a^3}\sum_{p,q}X_p^0X_q^0\sin{\frac{p\pi }{N}}\sin{\frac{q\pi}{N}}(\delta_{p,q-k}+\delta_{p,q+k})
\end{split}
\label{23c}
\end{equation}

%\twocolumngrid 

 Using \cref{23c} in \cref{23} and we arrive at the dynamical
 equation for $X_p^1$
%\pagebreak
\begin{equation}
    \begin{split}
        &\frac{\partial X_p^1}{\partial t} =-\frac{2\nu}{a^2}\Big(1-\cos{\frac{p\pi }{N}}\Big) X_p^1\\
         &+\frac{1}{4a^2}\sum_{p}\sum_q X_p^0X_q^0\sin{\frac{p\pi }{N}}\sin{\frac{q\pi}{N}}[\delta_{p,q+k}+\delta_{p,q-k}]\\
        &=-k_p X_p^1+\frac{1}{4a^2}\sum_{q}X_{p+q}^0X_q^0\sin{\frac{(q+p)\pi }{N}}\sin{\frac{q\pi}{N}}\\
        &+\frac{1}{4a^2}\sum_{q}X_{q-p}^0X_q^0\sin{\frac{(q-p)\pi }{N}}\sin{\frac{q\pi}{N}}
    \end{split}
    \label{25}
\end{equation}
The general solution is
\begin{equation}
    \begin{split}
        X_p^1(t)&=\frac{1}{4a^2}\sum\int_{0}^{t}dt^\prime e^{-k_p(t-t^\prime)}\Big[X_{p+q}^0(t^\prime)X_q^0(t^\prime)\sin{\frac{(p+q)\pi}{N}}\\
        &\sin{\frac{q\pi}{N}}+X_{q-p}^0(t^\prime)X_q^0(t^\prime)\sin{\frac{(q-p)\pi}{N}}\sin{\frac{q\pi}{N}}\Big]
    \end{split}
    \label{26}
\end{equation}
with the initial condition $X_p^1(0)=0$.

Using the general solution in \cref{26} the detail calculations of
two-time correlation function in the order $\lambda^2$ has been
presented in the Appendix(\ref{111}). Combining
Eqs.(\ref{eq:correlation_two_time_EW}), (\ref{19}) and (\ref{A4}), we get
%\begin{equation}
   % \begin{split}
   % \langle X_p^1(t_1)X_p^1(t_2)\rangle&=\frac{1}{64\nu^2}\Big[2Dt_1-\frac{DL^2}{\nu\pi^2}\Big(1-e^{-\frac{2\nu\pi^2 t_1}{L^2}}\Big)\Big]\\
  %  &\Big[2Dt_2-\frac{DL^2}{\nu\pi^2}\Big(1-e^{-\frac{2\nu\pi^2 t_2}{L^2}}\Big)\Big]
    %\end{split}
   % \label{27}
%\end{equation}
%Detailed calculation of Eq.(\ref{27}) is given in the Appendix(\ref{111}).

%So using all the values of the correlations from Eq.(\ref{18}) and Eq.(\ref{A4}) we get Eq.(\ref{19}) as follows

%\begin{equation}
 %   \begin{split}
  %      &\langle h_0(t_1)h_0(t_2)\rangle\\
   %     &=2Dt_2+2D\frac{L^2}{2\nu\pi^2}\Big(e^{-\frac{\nu\pi^2(t_1-t_2)}{L^2}}-e^{-\frac{\nu\pi^2(t_1+t_2)}{L^2}}\Big)\\
    %    &+\frac{D^2\lambda^2}{4\nu^2}\Big[t_1-\frac{L^2}{2\nu\pi^2}(1-e^{-\frac{2\nu\pi^2 t_1}{L^2}})\Big]\Big[t_2-\frac{L^2}{2\nu\pi^2}(1-e^{-\frac{2\nu\pi^2 t_2}{L^2}})\Big]
    %\end{split}
    %\label{28}
%\end{equation}

\begin{equation}
\begin{split}
&\langle h_0(t_1)h_0(t_2)\rangle\\
&=\frac{D}{L}t_2+\frac{D}{L}\frac{L^2}{\nu\pi^2}\Big[e^{-\frac{\nu\pi^2}{L^2}(t_1-t_2)}-e^{-\frac{\nu\pi^2}{L^2}(t_1+t_2)}\Big]\\
&+\frac{D^2}{L^2}\frac{\lambda^2}{16\nu^2}\Big[t_1-\frac{L^2}{2\nu\pi^2}(1-e^{-\frac{2\nu\pi^2 t_1}{L^2}})\Big]\Big[t_2-\frac{L^2}{2\nu\pi^2}(1-e^{-\frac{2\nu\pi^2 t_2}{L^2}})\Big]
\end{split}
\label{28}
\end{equation}

\subsection{For a finite value of $t$ but $L\to 0$}
In the finite $t$ domain if $L\rightarrow 0$ the term related to $L^2$ can be neglected. Putting this condition in Eq.(\ref{28}),
\begin{equation}
  \begin{split}
    \langle h_0(t_1)h_0(t_2)\rangle_{L\rightarrow
      0}&=\frac{D}{L}t_2+\frac{D^2}{L^2}\frac{\lambda^2}{16\nu^2}t_1t_2\\
    &=\frac{D}{L}t_2\left[1+\frac{D}{L}\frac{\lambda^2
    }{16\nu^2}t_1\right]
    \end{split}
    \label{29}
\end{equation}

Let us take spatial transformation as
$H(t)=\frac{h_0(t)}{\sqrt{\langle h_0^2(t)\rangle}}$ and we get
\begin{equation}
    \begin{split}
      \langle H(t_1)H(t_2)\rangle&=\frac{\langle
        h_0(t_1)h_0(t_2)\rangle}
      {\sqrt{\langle h_0^2(t_1)\rangle\langle h_0^2(t_2)\rangle}}\\
      &=\sqrt{\frac{\frac{D}{L}t_2}{\frac{D}{L}t_1}}\sqrt{\frac{1+\frac{D}{L}\frac{\lambda^2
          }{16\nu^2}t_1}{1+\frac{D}{L}\frac{\lambda^2 }{16\nu^2}t_2}}
    \end{split}
    \label{30}
\end{equation}

\begin{figure}
  \centering
  \includegraphics[width=0.50\textwidth]{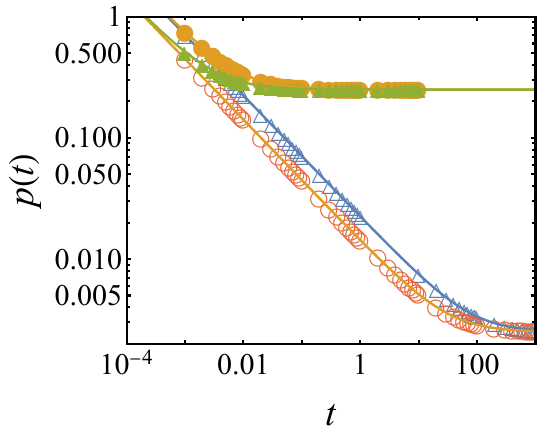}
  \caption{Log-Log plot of $p(t)$ with time $t$ for different values
    of $L$: $L=0.05$ (open and filled circles) and $L=0.02$
    (open and filled triangles) and for two choices for the
    values of $\nu$ and $\lambda$:$\lambda=0.01$ and $\nu=0.1$ (open
    symbols) and $\lambda=0.1$ and $\nu=0.01$ (filled
    symbols). The solid lines are the representations of
    Eq.(\ref{31}).}
    \label{fig:fig21}
\end{figure}

Now Time transformation may be taken as
$e^T=\frac{t}{1+\frac{D}{L}\frac{\lambda^2t}{16\nu^2}}$.

This gives, $\langle H(T_1)H(T_2)\rangle=e^{-(T_1-T_2)/2}$ Following
Slepian\cite{slepian1962}, if the correlation function of a
stochastic variable decays exponentially for all times
$C(T)=e^{-\lambda T}$, then the persistence probability is given
by
\begin{equation}
    P(T)=\frac{2}{\pi}\sin^{-1}(e^{-\lambda T})
\end{equation}
Asymptotically, $P(T)$ takes the form $P(T)\sim e^{-\lambda
  T}$. Consequently, in real time the persistence probability is found
as
\begin{equation}
    p(t)_{L\rightarrow 0}\sim\sqrt{\frac{1}{\frac{D}{L}t}+\frac{\lambda^2}{16\nu^2}}
    \label{31}
\end{equation}
It is quite interesting that the expression of $p(t)$ in the
asymptotic limit of $t\to\infty$ goes to a constant value of
$\lambda^2/16\nu^2$. In principle, one can use this result to extract
the ratio of $\lambda/\nu$, with the advantage being that the system
size required to extract the information need not be very large.

Fig.(\ref{fig:fig21}) shows the simulation results of $p(t)$
with respect to time $t$ for the condition $L\rightarrow 0$, which
actually validates the analytical expression of Eq.(\ref{31}). The
simulation is carried out taking very small values of $L$ for
different set of $\nu$ and $\lambda$ values. It is quite clearly seen
that at long-time limit $p(t)$ goes constant as found in the
analytical expression Eq.(\ref{31}). The simulation has been done
using the discrete form of the equation in Eq.(\ref{8}). The
trajectories were evolved in time with an integration time step of
$\delta t=0.001$. A total of $10^8$ trajectories were used in
estimating the survival probability.

\section{Conclusion}
In conclusion, we have investigated the persistence probability in
models of surface growth which are restricted by a finite domain, in
particular we have determined the expression for the persistence
probability $p(t)$ for two models of surface growth: the linear model
of Edwards-Wilkinson and that of the non-linear model of
Kardar-Parisi-Zhang. Our aim was to see whether a finite size of the
domain can help us in determining the physical parameters of the
equations - that is $\nu$ and $\lambda$ (see \cref{EW} and \cref{7}). 
In the case of the Edwards-Wilkinson model, the value of $\nu$ can be
determined from the decay of the quantity $t^{1/2} p(t)$ in the limit
of $t <<L^2/\nu\pi^2$ while for the nonlinear equation we find that in the
limit of $L \to 0$, we can extract the ratio $\nu /\lambda$ from
$p(t)$. The advantage of this study is that the domain size need not
be large to estimate these parameters and therefore can be efficiently
used in numerical simulations as well as experiments.
%\section{Acknowledgement}
% \section{DECLARATION OF INTERESTS}
% The authors declare no competing interests.
\bibliographystyle{unsrt}

\cleardoublepage

\stepcounter{myequation}
\stepcounter{myfigure}

\onecolumngrid

%\vspace{10pt}

% \vspace{1pt}

%\newpage

\begin{center}
{\bf {\Large{Appendix}}}
\end{center}
\begin{appendix}
\section{Calculation of $\langle X_p^1(t_1)X_q^1(t_2)\rangle$}
\label{111}
In this appendix we show the detailed calculation of the order
$\lambda^2$ term in the two-time correlation $\langle h_0(t_1)h_0(t_2)\rangle$. 
From \cref{19} the order $\lambda^2$ term reads as $\sum_p \sum_q
\langle X_p^1(t_1) X_q^1(t_2) \rangle$.
% \begin{equation}
%   \label{eq:order_lambdasq}
  
% \end{equation}

\begin{equation}
    \begin{split}
      &\langle
      X_p^1(t_1)X_q^1(t_2)\rangle=\frac{1}{16a^4}\Big[\int_{0}^{t_1}dt_1^\prime\int_{0}^{t_2}dt_2^\prime\sum_{p_1,q_1}\langle
      X_{p+p_1}^0(t_1^\prime)
      X_{p_1}^0(t_1^\prime)X_{q+q_1}^0(t_2^\prime)X_{q_1}^0(t_2^\prime)\rangle e^{-k_p(t_1-t_1^\prime)}e^{-k_q(t_2-t_2^\prime)}\sin{\frac{(p+p_1)\pi}{N}}\sin{\frac{p_1\pi}{N}}\\
      &\sin{\frac{(q+q_1)\pi}{N}}\sin{\frac{q_1\pi}{N}}+\int_{0}^{t_1}dt_1^\prime\int_{0}^{t_2}dt_2^\prime\sum_{p_1,q_1}\langle
      X_{p+p_1}^0(t_1^\prime)X_{p_1}^0(t_1^\prime)
      X_{q_1-q}^0(t_2^\prime)X_{q_1}^0(t_2^\prime)\rangle  e^{-k_p(t_1-t_1^\prime)}e^{-k_q(t_2-t_2^\prime)}\sin{\frac{(p+p_1)\pi}{N}}\sin{\frac{p_1\pi}{N}}\\
      &\sin{\frac{(q_1-q)\pi}{N}}\sin{\frac{q_1\pi}{N}}
      +\int_{0}^{t_1}dt_1^\prime\int_{0}^{t_2}dt_2^\prime\sum_{p_1,q_1}\langle
      X_{p_1-p}^0(t_1^\prime)X_{p_1}^0(t_1^\prime)X_{q+q_1}^0(t_2^\prime)X_{q_1}^0(t_2^\prime)\rangle
      e^{-k_p(t_1-t_1^\prime)}e^{-k_q(t_2-t_2^\prime)}\sin{\frac{(p_1-p)\pi}{N}}\sin{\frac{p_1\pi}{N}}\\
      &\sin{\frac{(q+q_1)\pi}{N}}\sin{\frac{q_1\pi}{N}}+\int_{0}^{t_1}dt_1^\prime\int_{0}^{t_2}dt_2^\prime\sum_{p_1,q_1}\langle
      X_{p_1-p}^0(t_1^\prime)X_{p_1}^0(t_1^\prime)
      X_{q_1-q}^0(t_2^\prime)X_{q_1}^0(t_2^\prime)\rangle e^{-k_p(t_1-t_1^\prime)}e^{-k_q(t_2-t_2^\prime)}\sin{\frac{(p_1-p)\pi}{N}}\sin{\frac{p_1\pi}{N}}\\
      &\sin{\frac{(q_1-q)\pi}{N}}\sin{\frac{q_1\pi}{N}}\Big]
    \end{split}
    \label{A1}
  \end{equation}
Since $X_p^0$ is a Gaussian stochastic process, the correlation
function $\langle
X_{p+p_1}^0(t_1^\prime)X_{p_1}^0(t_1^\prime)X_{q+q_1}^0(t_2^\prime)X_{q_1}^0(t_2^\prime)\rangle$
can be decomposed as
\begin{equation}
    \begin{split}
      &\sum_{p_1,q_1}\langle
      X_{p+p_1}^0(t_1^\prime)X_{p_1}^0(t_1^\prime)X_{q+q_1}^0(t_2^\prime)X_{q_1}^0(t_2^\prime)\rangle\sin{\frac{(p+p_1)\pi}{N}}\sin{\frac{p_1\pi}{N}}
      \sin{\frac{(q+q_1)\pi}{N}}\sin{\frac{q_1\pi}{N}}\\
      &=\sum_{p_1,q_1}\left[\langle
        X_{p+p_1}^0(t_1^\prime)X_{p_1}^0(t_1^\prime)\rangle\langle
        X_{q+q_1}^0(t_2^\prime)X_{q_1}^0(t_2^\prime)\rangle+\langle
        X_{p+p_1}^0(t_1^\prime) X_{q+q_1}^0(t_2^\prime)\rangle\langle
        X_{p_1}^0(t_1^\prime)X_{q_1}^0(t_2^\prime)\rangle \right.\\
      &\qquad\qquad\qquad\qquad\left. +\langle
        X_{p+p_1}^0(t_1^\prime)X_{q_1}^0(t_2^\prime)\rangle\langle
        X_{p_1}^0(t_1^\prime)X_{q+q_1}^0(t_2^\prime)\rangle\right]\sin{\frac{(p+p_1)\pi}{N}}\sin{\frac{p_1\pi}{N}}
      \sin{\frac{(q+q_1)\pi}{N}}\sin{\frac{q_1\pi}{N}}\\
      &=\frac{D^2}{4L^2}\Big[\mathlarger{\mathlarger{\sum}_{p_1,q_1}}\delta_{p+p_1,p_1}\frac{1-e^{-k_{p+p_1}t_1^\prime}e^{-k_{p_1}t_1^\prime}}{k_{p_1}+k_{p+p_1}}
      \delta_{q+q_1,q_1}
      \frac{1-e^{-k_{q+q_1}t_2^\prime}e^{-k_{q_1}t_2^\prime}}{k_{q+q_1}+k_{q_1}}\sin{\frac{(p+p_1)\pi}{N}}\sin{\frac{p_1\pi}{N}}
      \sin{\frac{(q+q_1)\pi}{N}}\sin{\frac{q_1\pi}{N}}\\
      &+\mathlarger{\mathlarger{\sum}_{p_1,q_1}}\delta_{p+p_1,q+q_1}
      \frac{e^{-k_{p+p_1}|t_1^\prime-t_2^\prime|}-e^{-k_{q+q_1}t_2^\prime}
        e^{-k_{p+p_1}t_1^\prime}}{k_{p+p_1}+k_{q+q_1}}\delta_{p_1,q_1}\frac{e^{-k_{p_1}|t_1^\prime-t_2^\prime|}-e^{-k_{q_1}t_2^\prime}e^{-k_{p_1}t_1^\prime}}{k_{p_1}+k_{q_1}}
      \sin{\frac{(p+p_1)\pi}{N}}\sin{\frac{p_1\pi}{N}}\sin{\frac{(q+q_1)\pi}{N}}\sin{\frac{q_1\pi}{N}}\\
      &+\mathlarger{\mathlarger{\sum}_{p_1,q_1}}\delta_{p+p_1,q_1}
      \frac{e^{-k_{p+p_1}|t_1^\prime-t_2^\prime|}
        -e^{-k_{q_1}t_2^\prime}e^{-k_{p+p_1}t_1^\prime}}{k_{p+p_1}+k_{q_1}}\delta_{p_1,q+q_1}\frac{e^{-k_{p_1}|t_1^\prime-t_2^\prime|}-e^{-k_{q+q_1}t_2^\prime}
        e^{-k_{p_1}t_1^\prime}}{k_{p_1}+k_{q+q_1}}\Big]\sin{\frac{(p+p_1)\pi}{N}}\sin{\frac{p_1\pi}{N}}
      \sin{\frac{(q+q_1)\pi}{N}}\sin{\frac{q_1\pi}{N}}\\
      &=\frac{D^2}{4L^2}\Big[\mathlarger{\mathlarger{\sum}_{p_1,q_1}}
      \delta_{p,0}\delta_{q,0}
      \frac{1-e^{-k_{p+p_1}t_1^\prime}e^{-k_{p_1}t_1^\prime}}{k_{p_1}+k_{p+p_1}}\frac{1-e^{-k_{q+q_1}t_2^\prime}e^{-k_{q_1}t_2^\prime}}{k_{q+q_1}+k_{q_1}}
      \sin{\frac{(p+p_1)\pi}{N}}\sin{\frac{p_1\pi}{N}}\sin{\frac{(q+q_1)\pi}{N}}\sin{\frac{q_1\pi}{N}}\\
      &+\mathlarger{\mathlarger{\sum}_{p_1}}\delta_{p,q}\frac{e^{-k_{p+p_1}|t_1^\prime-t_2^\prime|}
        -e^{-k_{q+p_1}t_2^\prime}e^{-k_{p+p_1}t_1^\prime}}{k_{p+p_1}+k_{q+p_1}}
      \frac{e^{-k_{p_1}|t_1^\prime-t_2^\prime|}-
        e^{-k_{p_1}(t_1^\prime+t_2^\prime)}}{2k_{p_1}}\sin{\frac{(p+p_1)\pi}{N}}\sin{\frac{(q+p_1)\pi}{N}}\sin^2{\frac{p_1\pi}{N}}\\
      &+\mathlarger{\mathlarger{\sum}_{p_1}}\delta_{p,-q}\frac{e^{-k_{p+p_1}|t_1^\prime-t_2^\prime|}
        -e^{-k_{p+p_1}(t_1^\prime+t_2^\prime)}}{2k_{p+p_1}}
      \frac{e^{-k_{p_1}|t_1^\prime-t_2^\prime|}-e^{-k_{p_1}t_1^\prime}e^{-k_{p+q+p_1}t_2^\prime}}{k_{p_1}+k_{p+q+p_1}}\sin^2{\frac{(p+p_1)\pi}{N}}\sin{\frac{p_1\pi}{N}}
      \sin{\frac{(p+q+p_1)\pi}{N}}\Big]
    \end{split}
    \label{A2}
  \end{equation}
  Before we carefully examine \cref{A2} term by term, we note that in
  the two-time correlation function in \cref{19}, the term in the
  order of $\lambda^2$ has a double sum over the Fourier modes denoted
  by $p$ and $q$. Consequently, in the first term in \cref{A2}, this
  double sum picks up the modes $p=0$ and $q=0$ and therefore, even
  for the choice of the lowest value of $p_1=1$, the first term
  corresponds to largest time scale $\tau^{-1}_1=\nu \pi^2/L^2$. In
  contrast, when we look at the second and the third term, the first
  term in the sum corresponds to $p=1,p_1=1$ with
  $p+p_1=2$. Therefore, the smallest relaxation time scale that appears
  in these terms correpsond to $\tau^{-1}_4=4\nu
  \pi^2/L^2$. Consequently, in our final expression we ignore the two terms.
  
Looking at the three other terms in \cref{A1}, the four-point
correlation function can be similarly decomposed as product of
two-point correlation functions.
\begin{equation}
  \label{eq:term2}
    \begin{split}
      &\sum_{p_1,q_1}\langle
      X_{p+p_1}^0(t_1^\prime)X_{p_1}^0(t_1^\prime)X_{q_1-q}^0(t_2^\prime)X_{q_1}^0(t_2^\prime)\rangle\sin{\frac{(p+p_1)\pi}{N}}\sin{\frac{p_1\pi}{N}}
      \sin{\frac{(q_1-q)\pi}{N}}\sin{\frac{q_1\pi}{N}}\\
      &=\sum_{p_1,q_1}\left[\langle
        X_{p+p_1}^0(t_1^\prime)X_{p_1}^0(t_1^\prime)\rangle\langle
        X_{q_1-q}^0(t_2^\prime)X_{q_1}^0(t_2^\prime)\rangle+\langle
        X_{p+p_1}^0(t_1^\prime) X_{q_1-q}^0(t_2^\prime)\rangle\langle
        X_{p_1}^0(t_1^\prime)X_{q_1}^0(t_2^\prime)\rangle \right.\\
      &\qquad\qquad\qquad\qquad\left. +\langle
        X_{p+p_1}^0(t_1^\prime)X_{q_1}^0(t_2^\prime)\rangle\langle
        X_{p_1}^0(t_1^\prime)X_{q_1-q}^0(t_2^\prime)\rangle\right]\sin{\frac{(p+p_1)\pi}{N}}\sin{\frac{p_1\pi}{N}}
      \sin{\frac{(q_1-q)\pi}{N}}\sin{\frac{q_1\pi}{N}}\\
\end{split}
\end{equation}
\begin{equation}
  \label{eq:term3}
      \begin{split}
      &\sum_{p_1,q_1}\langle
      X_{p_1-p}^0(t_1^\prime)X_{p_1}^0(t_1^\prime)X_{q+q_1}^0(t_2^\prime)X_{q_1}^0(t_2^\prime)\rangle\sin{\frac{(p_1-p)\pi}{N}}\sin{\frac{p_1\pi}{N}}
      \sin{\frac{(q+q_1)\pi}{N}}\sin{\frac{q_1\pi}{N}}\\
      &=\sum_{p_1,q_1}\left[\langle
        X_{p_1-p}^0(t_1^\prime)X_{p_1}^0(t_1^\prime)\rangle\langle
        X_{q+q_1}^0(t_2^\prime)X_{q_1}^0(t_2^\prime)\rangle+\langle
        X_{p_1-p}^0(t_1^\prime) X_{q+q_1}^0(t_2^\prime)\rangle\langle
        X_{p_1}^0(t_1^\prime)X_{q_1}^0(t_2^\prime)\rangle \right.\\
      &\qquad\qquad\qquad\qquad\left. +\langle
        X_{p_1-p}^0(t_1^\prime)X_{q_1}^0(t_2^\prime)\rangle\langle
        X_{p_1}^0(t_1^\prime)X_{q_1+q}^0(t_2^\prime)\rangle\right]\sin{\frac{(p_1-p)\pi}{N}}\sin{\frac{p_1\pi}{N}}
      \sin{\frac{(q+q_1)\pi}{N}}\sin{\frac{q_1\pi}{N}}\\
\end{split}
\end{equation}

\begin{equation}
  \label{eq:term4}
      \begin{split}
      &\sum_{p_1,q_1}\langle
      X_{p_1-p}^0(t_1^\prime)X_{p_1}^0(t_1^\prime)X_{q_1-q}^0(t_2^\prime)X_{q_1}^0(t_2^\prime)\rangle\sin{\frac{(p_1-p)\pi}{N}}\sin{\frac{p_1\pi}{N}}
      \sin{\frac{(q_1-q)\pi}{N}}\sin{\frac{q_1\pi}{N}}\\
      &=\sum_{p_1,q_1}\left[\langle
        X_{p_1-p}^0(t_1^\prime)X_{p_1}^0(t_1^\prime)\rangle\langle
        X_{q_1-q}^0(t_2^\prime)X_{q_1}^0(t_2^\prime)\rangle+\langle
        X_{p_1-p}^0(t_1^\prime) X_{q_1-q}^0(t_2^\prime)\rangle\langle
        X_{p_1}^0(t_1^\prime)X_{q_1}^0(t_2^\prime)\rangle \right.\\
      &\qquad\qquad\qquad\qquad\left. +\langle
        X_{p_1-p}^0(t_1^\prime)X_{q_1}^0(t_2^\prime)\rangle\langle
        X_{p_1}^0(t_1^\prime)X_{q_1-q}^0(t_2^\prime)\rangle\right]\sin{\frac{(p_1-p)\pi}{N}}\sin{\frac{p_1\pi}{N}}
      \sin{\frac{(q_1-q)\pi}{N}}\sin{\frac{q_1\pi}{N}}\\
\end{split}
\end{equation}

The first terms in the all the three expression will have
$\delta_{p,0}\delta_{q,0}$ and therefore we retain these terms in the
final expression of $\langle X_P^1(t_1)X_q^1(t_2)\rangle$. 
% we neglect second and third terms, only take first
% one and put it into the Eq.(\ref{A1}). In Eq.(\ref{A1}) second, third
% and fourth terms are zero. For those terms when we break the summation
% and take $p=1$, $p_1=1$, $q=1$ and $q_1=1$, we get
% $\sin{\frac{(q_1-q)\pi}{N}}=0$(for second term),
% $\sin{\frac{(p_1-p)\pi}{N}}=0$ (for third term), and
% $\sin{\frac{(q_1-q)\pi}{N}}=0$(for the fourth term).So finally we
% get,
We have, after carrying over the sum over $p$ and $q$
\begin{equation}
    \begin{split}
    \sum_p \sum_q \langle
    X_p^1(t_1)X_q^1(t_2)\rangle&=\frac{4 D^2}{4L^2}\frac{1}{16a^4}\int_{0}^{t_1^\prime}dt_1^\prime\int_{0}^{t_2^\prime}dt_2^\prime\mathlarger{\mathlarger{\sum}_{p_1}}\frac{1-e^{-2k_{p_1}t_1^\prime}}{2k_{p_1}}
    \mathlarger{\mathlarger{\sum}_{q_1}}\frac{1-e^{-2k_{q_1}t_2^\prime}}{2k_{q_1}}\sin^2{\frac{p_1\pi}{N}}\sin^2{\frac{q_1\pi}{N}}
    \end{split}
\label{A3}
\end{equation}
To break the summation, putting $p_1=1, q_1=1$, we get $\sin{\frac{p_1\pi}{N}}\sim\frac{p_1\pi}{N}$, for $N\rightarrow\infty$, So Eq.(\ref{A3}) becomes

\begin{equation}
    \begin{split}
         \sum_p \sum_q \langle X_p^1(t_1)X_q^1(t_2)\rangle&=\frac{D^2}{L^2}\frac{1}{16a^4}\int_{0}^{t_1^\prime}dt_1^\prime\frac{1-e^{-\frac{2\nu\pi^2t_1^\prime}{L^2}}}
         {\frac{2\nu\pi^2}{L^2}}\left(\frac{\pi^2
             a^2}{L^2}\right)\int_{0}^{t_2^\prime}dt_2^\prime\frac{1-e^{-\frac{2\nu\pi^2t_2^\prime}{L^2}}}{\frac{2\nu\pi^2}{L^2}}
         \Big(\frac{\pi^2a^2}{L^2}\Big)\\
        &=\frac{D^2}{L^2}\frac{1}{64\nu^2}\Big[t_1-\frac{L^2}{2\nu\pi^2}\Big(1-e^{-\frac{2\nu\pi^2
            t_1}{L^2}}\Big)\Big]\Big[t_2-\frac{L^2}{2\nu\pi^2}\Big(1-
        e^{-\frac{2\nu\pi^2 t_2}{L^2}}\Big)\Big]
    \end{split}
    \label{A4}
\end{equation}

\section{Calculation for finite $L$ value}
\label{1111}
In this section we provide the detailed calculation of the two-time
correlation function $\langle h_0(t_1) h_0(t_2)$. Starting frrom Eq.(\ref{28}) we get
\begin{equation}
    \begin{split}
        &\langle h_0(t_1)h_0(t_2)\rangle=\langle X_0^0(t_1)X_0^0(t_2)\rangle+4\sum_{p,q}\langle X_p^0(t_1)X_q^0(t_2)\rangle
    +4\lambda^2\sum_{p,q}\langle X_p^1(t_1)X_q^1(t_2)\rangle\\
        &=\frac{D}{L}t_2+\frac{2D}{L}\frac{L^2}{2\nu\pi^2}\Big(e^{-\frac{\nu\pi^2(t_1-t_2)}{L^2}}-e^{-\frac{\nu\pi^2(t_1+t_2)}{L^2}}\Big)
        +\frac{D^2\lambda^2}{16\nu^2L^2}\Big[t_1-\frac{L^2}{2\nu\pi^2}(1-e^{-\frac{2\nu\pi^2
            t_1}{L^2}})\Big]\Big[t_2-\frac{L^2}{2\nu\pi^2}
        (1-e^{-\frac{2\nu\pi^2 t_2}{L^2}})\Big]\\
        &=\frac{D}{L}t_2\Bigg[1+\Big(\frac{L^2}{\nu\pi^2t_2}\Big)\Big(e^{-\frac{\nu\pi^2(t_1-t_2)}{L^2}}-e^{-\frac{\nu\pi^2(t_1+t_2)}{L^2}}\Big)\Bigg]
        +\frac{D^2\lambda^2}{16\nu^2L^2}
        \Big[t_1-\frac{L^2}{2\nu\pi^2}(1-e^{-\frac{2\nu\pi^2 t_1}{L^2}})\Big]\Big[t_2-\frac{L^2}{2\nu\pi^2}(1-e^{-\frac{2\nu\pi^2 t_2}{L^2}})\Big]\\
        &=\frac{D}{L}t_2\Bigg[1+\Big(\frac{L^2}{\nu\pi^2t_2}\Big)\Big(e^{-\frac{\nu\pi^2(t_1-t_2)}{L^2}}-e^{-\frac{\nu\pi^2(t_1+t_2)}{L^2}}\Big)\Bigg]
        +\frac{D^2\lambda^2}{16\nu^2L^2}t_1t_2
        \Big[1-\frac{L^2}{2\nu\pi^2t_1}(1-e^{-\frac{2\nu\pi^2
            t_1}{L^2}})\Big]\Big[1-\frac{L^2}{2\nu\pi^2t_2}(1-e^{-\frac{2\nu\pi^2
            t_2}{L^2}})\Big]\\
        &=\frac{D}{L}t_2\Bigg[1+\Big(\frac{2L^2}{\nu\pi^2t_2}\Big)e^{-\frac{\nu
        \pi^2}{L^2} t_1}\sinh \frac{\nu \pi^2 t_2}{L^2}\Bigg]
        +\frac{D^2\lambda^2}{16\nu^2L^2}t_1t_2
        \Big[1-\frac{L^2}{\nu\pi^2t_1}e^{-\frac{\nu\pi^2
            t_1}{L^2}}\sinh \frac{\nu \pi^2 t_1}{L^2}\Big]\Big[1-\frac{L^2}{\nu\pi^2t_2}e^{-\frac{\nu\pi^2
            t_2}{L^2}}\sinh \frac{\nu \pi^2 t_2}{L^2}\Big] \\
    \end{split}
    \label{A5}
\end{equation}

\end{appendix}

\pagebreak

\stepcounter{myequation}
\stepcounter{myfigure}

\renewcommand{\thefigure}{S\arabic{figure}}
\renewcommand{\theequation}{A\arabic{equation}}

\onecolumngrid

\vspace{10pt}

% \vspace{1pt}

\end{document}